# Downsampling of optical frequency combs

DANIEL C. COLE,[1,2,*] SCOTT B. PAPP,[1,2] SCOTT A. DIDDAMS[1,2]

[1]National Institute of Standards and Technology (NIST), Boulder, CO 80305, USA
[2]Department of Physics, University of Colorado, Boulder, CO 80309, USA
*Corresponding author: daniel.cole@nist.gov

**We demonstrate repetition-rate downsampling of optical frequency combs by way of pulse gating. The reduced repetition rate enables increased pulse energy, facilitating efficient spectral broadening and f-2f interferometry. To explore the technique, we downsample a 250 MHz repetition-rate comb to 25 MHz and detect the carrier-envelope offset frequency of the downsampled pulse train. We investigate the effects of pulse gating on the noise properties of the pulse train and the limitations of the technique by characterizing the phase-noise spectrum of the downsampled comb and deliberately imposing timing jitter on the pulse gate. We show that, up to an expected reduction modulo the new repetition rate, downsampling neither shifts nor introduces noise to the carrier-envelope offset frequency of the comb above the level of several microhertz. Additionally, we discuss the effect of downsampling on the spectrum of intensity fluctuations of the optical pulse train. Finally, we discuss some practical considerations relevant for the application of the technique to frequency combs with repetition rates in the 10 GHz range and higher.**

## 1. INTRODUCTION

The optical frequency comb is a critical tool in the measurement of optical frequencies, enabling applications including optical clocks, optical waveform generation, precision spectroscopy, and low-noise frequency synthesis[1,2]. Well-established ultrafast laser technology yields commercial-grade frequency combs with repetition rates typically ranging from 0.1 to 1 GHz, and stabilization of mode-locked laser spectra has been achieved at repetition rates up to 10 GHz under well-controlled conditions in a research setting[3]. However, at the highest repetition rates, frequency comb stabilization via *f-2f* self-referencing[4] is challenging due to the decreased pulse energy available for spectral broadening. This issue is more acute for emerging high-repetition rate comb platforms in low size, weight, and power packages. In particular, frequency combs generated in chip-integrable passive ring microresonators generate spectra with spacings between comb teeth in the range of 10 GHz to 1 THz[5–8]. While these combs have been self-referenced[9–13], their high repetition rates still present challenges in interfacing with more conventional electronics and in applications like spectroscopy, where finer spectral sampling might be desirable.

In this paper, we discuss repetition-rate downsampling of optical frequency combs to address these challenges and increase the versatility of high-repetition-rate comb systems for applications. Downsampling via pulse gating, also referred to as "pulse picking" in the literature, has proven useful in the context of high-field, phase-sensitive ultrafast optics for the generation of energetic, carrier-envelope-phase-stabilized ultrashort pulses[14,15]. In this application, a comb with initial repetition rate in the ~100 MHz range that has already been stabilized is pulse-picked to a repetition rate on the order of 1-100 kilohertz. Concerns in this application center around control and preservation of the carrier-envelope phase in the pulse-picking and amplification process[16,17]. In this paper we consider downsampling within the context of optical metrology with frequency combs, and we are concerned with downsampling's effect on the optical phase noise, the pulse-to-pulse energy fluctuations, and the carrier-envelope offset frequency. In particular, it is important that the downsampled pulse train is suitable for *f-2f* self-referencing.

The utility of repetition-rate downsampling in the spectral broadening of high-repetition-rate pulse trains has already been explored[18,19], and our group recently used downsampling in the first successful implementation of the technique to detect the carrier-envelope offset frequency of a 10 GHz comb[20]. Here, in order to provide a foundation for future applications of the technique, we report on our practical investigation into both the utility and limitations of repetition-rate downsampling in high-repetition-rate comb systems.

We use a commercial 250 MHz C-band Er:fiber frequency comb as a model system to demonstrate and investigate downsampling. We downsample the input comb to 25 MHz and count its offset frequency after amplification and spectral broadening (Fig 1). Using this system, we investigate the effects of downsampling on the properties of the optical pulse train in the presence of technical noise. We measure little change in the phase noise of the optical pulse train due to downsampling. Further, we demonstrate that in the presence of timing jitter on the electronic gating signal below a well-defined threshold, downsampling preserves the offset frequency of the comb to the level of several microhertz. We present a quantitative model for the fundamental impact of ideal downsampling, without technical limitations, on the pulse train's noise properties and compare it with experimental results in a 10 GHz comb system, which show that downsampling leads to a modest increase in pulse train's relative intensity noise (RIN) that is not prohibitive for metrology experiments. Finally, we model and numerically investigate the effects of finite and fluctuating extinction of rejected pulses during amplification and self-referencing of the downsampled pulse train, and we find that using an optical amplifier in the linear regime to amplify the downsampled pulse train minimizes negative effects of poor extinction.

## 2. RESULTS

### A. Demonstration of concept

The application of downsampling to the detection of the carrier-envelope offset frequency of a 250 MHz comb is shown in Fig. 1. Our pulse gating scheme, shown in Fig. 1a, employs a Mach-Zehnder (MZ) electro-optic intensity modulator driven by 25 MHz rectangular electronic gating pulses with 80 ps transitions and 3.5 ns duration. The electronic pulse generator and the repetition rate of the input 250 MHz comb are both referenced to a hydrogen maser to maintain synchronization. The DC bias of the EOM is set for maximum extinction outside the electronic gate, whose amplitude is matched to $V_\pi$ of the EOM. This downsampling scheme results in a stable 25 MHz optical pulse train with >12 dB contrast (Fig. 1b). This extinction ratio is sufficient for our investigation here but could be improved by cascading modulators with higher extinction ratios. The average power of our 250 MHz pulse train is reduced from 30 mW to 400 µW by the pulse gating process and the insertion loss of the optical components. We amplify the pulse train to 35 mW by use of a normal-dispersion erbium-doped fiber amplifier, which provides some spectral broadening and temporal pulse compression[21]. An octave-spanning supercontinuum is obtained by launching the amplified, <100 fs, ~1 nJ pulses into 20 cm of highly nonlinear fiber (HNLF)[22]; the resulting spectrum is shown in Fig. 1c. For comparison, we also present the supercontinuum generated by the 250 MHz comb with the EOM set for constant maximum transmission under otherwise identical conditions. The 250 MHz comb is amplified by the same EDFA to an average power of 85 mW, corresponding to 340 pJ pulse energy, before it enters the HNLF.

To detect $f_0$, the octave-spanning supercontinuum shown in Fig. 1c is sent into a free-space $f$-$2f$ interferometer consisting of a half-wave plate and a periodically poled lithium niobate (PPLN) crystal quasi-phase-matched for second-harmonic generation at 1980 nm. The generated 990 nm light is shown in Fig 1c. A 10 nm band-pass filter at 990 nm selects this second harmonic and the co-linear supercontinuum at 990 nm, which are then photodetected to observe $f_0$ with 30 dB signal-to-noise ratio, shown in Fig. 1d. Fig. 1e shows a 2000 s record of $f_0$ for the downsampled comb.

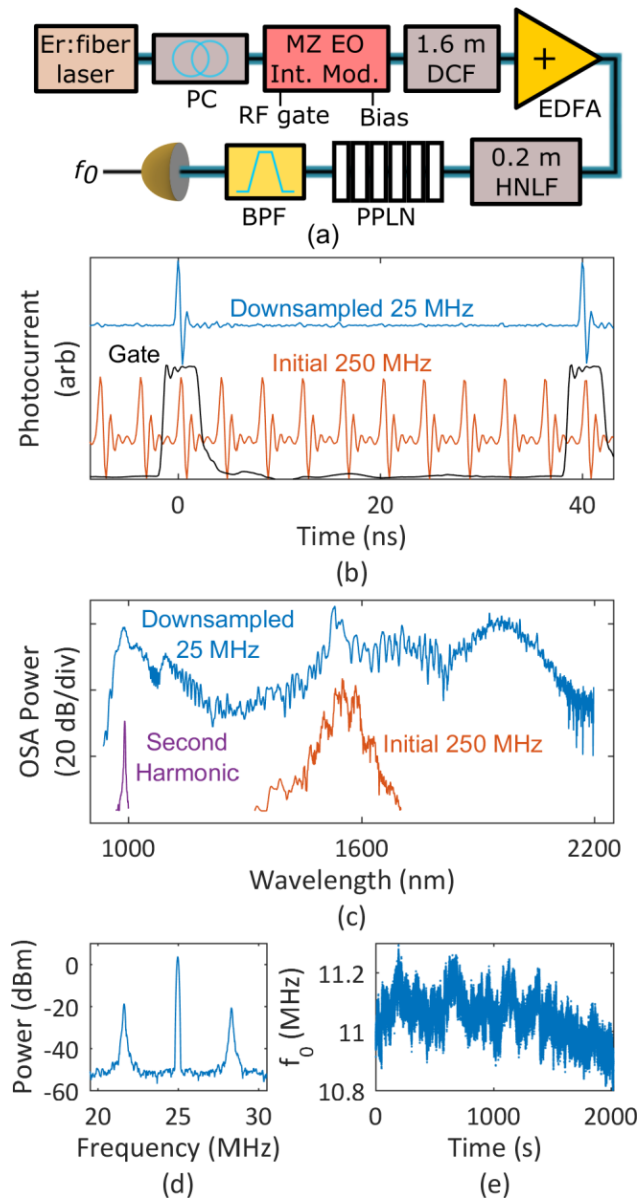

Figure 1. a) Schematic for downsampling a 250 MHz Er:fiber comb and detecting the offset frequency of the resulting 25 MHz pulse train. PC – polarization controller. DCF – dispersion-compensating fiber. EDFA – erbium-doped fiber amplifier. HNLF – highly nonlinear fiber. PPLN – periodically-poled lithium niobate. BPF – (optical) band-pass filter. b) Input (bottom, orange) and downsampled (top, blue) pulse trains from a 1 GHz photodetector, with the electronic gate superimposed on the input pulse train (black). c) Octave-spanning supercontinuum generated by downsampling (top, blue), second harmonic generated for $f_0$ detection (purple), and for comparison the supercontinuum generated by the same apparatus without downsampling (orange). d) Detected repetition rate and $f_0$ beat at 100 kHz RBW; signal-to-noise ratio of $f_0$ is 30 dB. e) Counted frequency of the detected free-running offset beat. Data is taken for ~2000 s at 10 ms gate time. The offset frequency of the 250 MHz commercial comb was adjusted between measurements shown in Figs. 1d and 1e to simplify electronic processing.

## B. Mathematical model for downsampling

While Fig. 1 presents an absolute frequency measurement of $f_0$ enabled by our downsampling technique, it does not demonstrate the deterministic connection between the input and downsampled combs that is essential for applications. To understand this relationship, we first present a simple model of downsampling, and then experimentally test its conclusions.

We model the gated pulse train's electric field as the product of the incoming comb's field and a time-varying amplitude modulation. For an incoming optical frequency comb with repetition rate $f_r$, complex single-pulse field $A(t)$ which is localized near $t = 0$, and pulse-to-pulse carrier-envelope phase shift $\phi$, pulse gating by a train of rectangular pulses of length $t_g$ and arrival rate $f_g$ yields a downsampled comb with field

$$a(t) = \left[\Sigma_n A(t - n/f_r) e^{in\phi}\right]\left[\Sigma_m \text{Rect}\left(\frac{t - m/f_g}{t_g}\right)\right] \quad (1)$$

where $\text{Rect}(x)$ is the rectangle function, taking the value 1 for $-1/2 \leq x \leq 1/2$ and 0 elsewhere. Indices $n$ and $m$ count the pulse number of the incoming pulse train and the electronic gate respectively. The optical spectrum of the downsampled pulse train $a(t)$, calculated via the convolution theorem for the Fourier transform, is:

$$\mathcal{F}\{a\}(f) \sim 4\pi f_r \sum_{nm} \frac{1}{m} \mathcal{F}\{A\}(f_0 + nf_r) \times \sin(\pi m t_g f_g) \delta(f - f_0 - nf_r - mf_g), \quad (2)$$

where $f_0 = f_r \cdot \phi/2\pi$ is the carrier-envelope offset frequency of the incoming comb. The downsampled pulse train has spectral content at optical modes $f_0 + nf_r$, as well as at intensity modulation sidebands whose frequency offsets $mf_g$ are harmonics of the gating frequency. To avoid the generation of unwanted modulations, pulse gating at an integer subharmonic of the incoming repetition rate, $f_g = f_r/N$, is essential. In this case superposition of the intensity modulation components created by pulse gating results in a downsampled frequency comb with a single mode spacing. Moreover, this model predicts that the offset frequency is preserved up to a reduction modulo the comb's new repetition rate.

Notably, for pulse gating at a sub-harmonic of the input comb's repetition rate, timing jitter of the electronic gate that is less than its duration does not contribute to noise on the downsampled comb. By modeling jitter as gate-to-gate arrival-time delays $\Delta t_m$, it can be shown that the downsampled comb's amplitude $a(t)$ and spectrum $\mathcal{F}\{a\}(f)$ do not deviate from Eqn. (1) provided that: 1) The jitter is a sufficiently small $|\Delta t_m| < t_g/2$, i.e., that the optical and electronic pulses are always substantially overlapped, and 2) That the optical pulses are substantially shorter than the electrical pulses, which is true for most systems. Thus, in general we expect that the carrier-envelope offset frequency of the incoming comb is preserved by downsampling even with jitter on the gate signal.

## C. Effect of technical noise in downsampling on a pulse train's noise properties

We supplement the mathematical model above with an experimental investigation of the effects of downsampling on the noise properties of the pulse train. First we consider the effects of technical limitations to ideal downsampling, and then we discuss fundamental effects associated with aliasing of high-Fourier-frequency optical noise and shot noise.

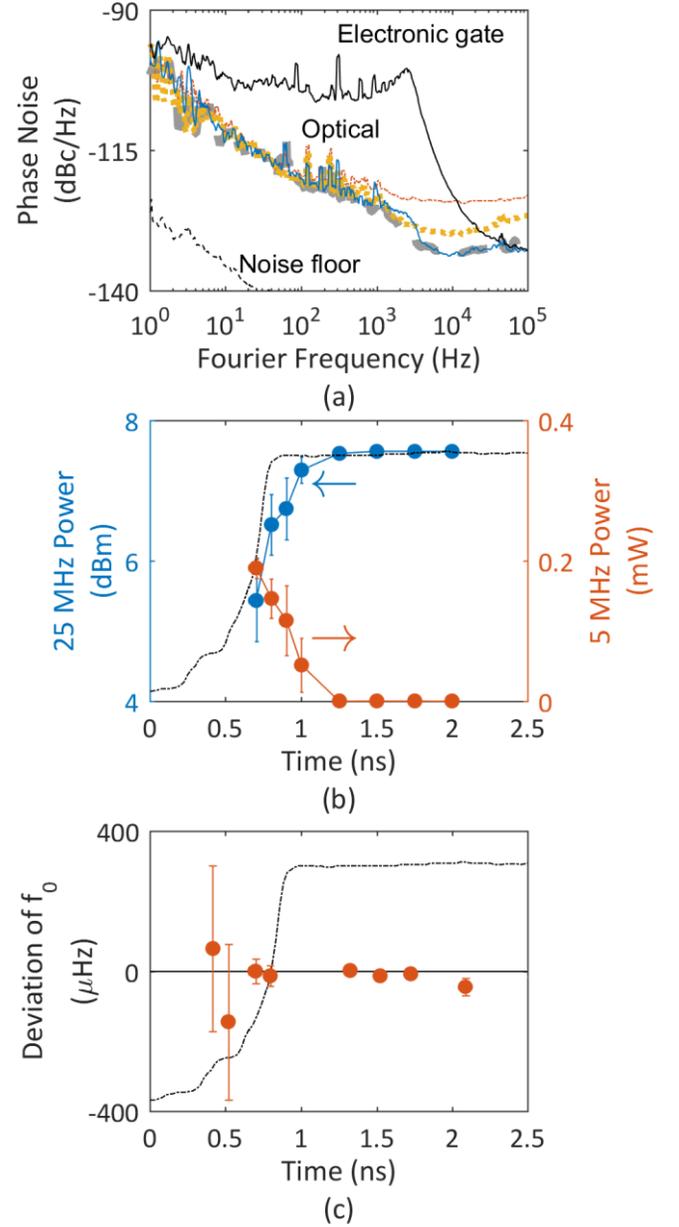

Figure 2. a) Measured repetition-rate phase noise of spectral components of the supercontinuum, selected by a 990±5 nm band pass filter (dot-dashed orange) 1650 nm long pass filter (dotted yellow) and the entire downsampled 25 MHz frequency comb measured immediately before the EDFA (solid blue), along with the 250 MHz comb (large-dashed gray, shifted by $20 \log(1/10) = -20$ dB). Also shown is the phase noise of the electronic gate generator (top, solid black). b) Amplitude of the downsampled pulse-train modulation due to 250 ps jitter at 5 MHz rate. The position of a data point on the x-axis indicates its mean position within the gate, shown in dashed black. Measurement uncertainties arise due to a latency between the optical trigger and the start of the electronic gating signal which varies on the order of 50 ps. c) Deviation of the carrier-envelope offset frequency of the downsampled comb from the 250 MHz comb's offset frequency as a function of the alignment of optical pulses within the gate.

We measure the phase-noise spectrum of the downsampled comb's repetition rate at different points in our apparatus, as shown in Fig. 2a. We also plot the phase noise of the 250 MHz comb, which has been shifted by $-10\log_{10} N^2 = -20$ dB to facilitate comparison[18], and the phase noise of the electronic gate. The downsampled frequency comb's phase-noise spectrum matches that of the 250 MHz comb except for a small increase at ~3 kHz, likely corresponding to the corner in the gate generator's phase noise at the same frequency. The phase noise of the high- and low-frequency ends of the supercontinuum similarly matches the 250 MHz comb below 1 kHz. The higher phase noise in the supercontinuum beyond 1 kHz is likely due to noise generation processes in the HNLF, such as the conversion of amplitude fluctuations on input pulses to timing jitter in the supercontinuum[23].

The timing jitter of our gating pulse train is between 5 ps (obtained by integrating the phase noise plotted in Fig. 2 to 100 kHz) and 10 ps (extrapolating constant phase noise to the 12.5 MHz Nyquist frequency and integrating). These jitter values are small relative to the 4 ns repetition period of the incoming optical pulse train. As the repetition rate of the incoming optical pulse train increases to >10 GHz, the gate duration must correspondingly decrease for single-pulse gating, and timing jitter on the gate may become a significant fraction of the gate duration. To explore the effects of timing jitter larger than our pulse generator's inherent 5 to 10 ps, we impose excess jitter on the gating signal. We modulate the relative timing between the gating signal and the incoming optical pulse train at a frequency of 5 MHz with an amplitude of 250 ps. The effect of this jitter is manifest in the microwave power of the gated comb as 5 MHz intensity-modulation sidebands whose amplitude depends on the position of the optical pulses within the gate, as shown in Fig. 2b. Pulses with a mean position within 250 ps of the gate edge are substantially modulated by the 5 MHz gate-delay signal. This agrees with the prediction of a sharp threshold on the acceptable level of timing jitter on the gate.

It is essential to establish that the comb's carrier-envelope offset frequency is preserved in the downsampling process. To do this, we perform a frequency comparison of the 25 MHz downsampled comb and a separate output of the 250 MHz comb. This 250 MHz output is intensity modulated so that a measurement of the nonzero optical heterodyne beat frequency between an intensity modulation sideband and a pulse-gating sideband of the downsampled comb reveals the relative frequency offset of the two combs. Figure 2c shows the null frequency shift between the 25 MHz and 250 MHz combs, which we have characterized for different alignments of the optical pulse within the gate. At the level of several microhertz, better than $10^{-18}$ relative to the 200 THz optical carrier frequency, we observe no frequency shift between the 250 MHz comb and the downsampled 25 MHz comb when the gate is properly aligned. This confirms the utility of downsampling for measurement of a high-repetition-rate comb's offset frequency for subsequent use of the comb in, for example, a spectroscopy experiment requiring high power per comb mode and high frequency precision.

**D. Fundamental effects of ideal downsampling on a pulse train's noise properties**

In addition to the conversion of electronic technical noise to optical noise on the downsampled pulse train, there exists a further mechanism by which downsampling can change the measured amplitude noise properties of the pulse train. Even ideal downsampling, free of electronic noise, leads to an increase in the measured power spectral density (PSD) of optical pulse energy fluctuations (PEF) when technical pulse energy noise is present. This is due to aliasing of components of the PSD of pulse energy fluctuations at frequencies above the Nyquist frequency, $f_r/2$, when the Nyquist frequency is reduced by downsampling. Assuming random fluctuations from pulse to pulse, downsampling does not change the RMS fractional pulse energy fluctuation $\sigma_{PEF}$, whose square is equal to the frequency integral of the PSD of pulse energy fluctuations $S_{PEF}(f)$:

$$\sigma_{PEF}^2 = \int_0^{\frac{f_r}{2}} df\, S_{PEF}(f). \quad (3)$$

Because the Nyquist frequency $f_r/2$ defines the upper limit for integration of $S_{PEF}$, in order for $\sigma_{PEF}$ to be preserved $S_{PEF}(f)$ must increase when the Nyquist frequency is reduced. For example, in the simple case of white technical noise on the pulse energies with density $S_o$, we have

$$\sigma_{PEF}^2 = \int_0^{f_r/2} df\, S_o = \int_0^{f_r/2N} df\, S', \quad (4)$$

which shows that downsampling must increase the measured PSD of white technical noise from $S_o$ to $S' = NS_o$, assuming there are no spectral correlations. However, this simple multiplicative increase is restricted to the case of white technical noise. In general, the PSD of pulse energy fluctuations of the new pulse train is determined from the original PSD through the usual method of modeling aliasing of a signal: a new Fourier frequency for each component of the original PSD is obtained by reducing the original Fourier frequency by a multiple of $-f_r/N$ so that it lies between $-f_r/2N$ and $f_r/2N$ and taking its absolute value. The new PSD is then determined by taking the quadrature sum of the PSD components at the same aliased Fourier frequency. This phenomenon is derived mathematically and demonstrated experimentally in Ref. 17, where the analysis of carrier-envelope phase noise applies equally well to pulse energy fluctuations.

In contrast with the increase in the PSD of pulse energy fluctuations arising from coincidence of the optical pulse with the edge of the electrical gate, which increases $\sigma_{PEF}$, the aliasing mechanism described above preserves $\sigma_{PEF}$. An important consequence of this is that while technical noise can lead to supercontinuum decoherence in external nonlinear spectral broadening, aliasing does not, because it is $\sigma_{PEF}$ which determines the degree of supercontinuum decoherence. Thus the aliasing mechanism impedes $f$-$2f$ self-referencing only by reducing the available signal-to-noise ratio of an $f_0$ signal in a straightforward linear fashion.

In practice, the relevance of the aliasing of the PSD of pulse energy fluctuations is determined by the presence of technical noise on the pulse energies at high Fourier frequency $f > f_r/2N$. For sufficiently small downsampling factors (e.g. $\frac{f_r}{2N} \geq \sim 50$ MHz) and depending on the comb source, it is possible that the only source of intensity noise at frequencies above $f_r/2N$ is shot noise. Shot noise results in a maximal (shot-noise-limited) signal-to-noise (SNR) ratio of an optical heterodyne beat with a local oscillator laser which is reduced by $N^2$ (in electrical power units) as the average power of the pulse train is reduced by downsampling by a factor of $N$. In contrast, in the case of detection of a carrier-envelope-offset beat with fixed optical detection bandwidth, the shot-noise-limited SNR is preserved in downsampling. One way to understand these results is to model the shot noise at a given Fourier frequency as the incoherent sum of optical heterodyne beats between each optical comb mode and the uncorrelated vacuum fluctuations at the appropriate optical frequency[24,25], and to take into account the fact that during downsampling the optical power of each comb mode is reduced by $N^2$, with the first factor of $N$ coming from reduction of the total optical power and the second factor of $N$ due to the increase in the spectral density of comb modes.

We experimentally investigate the impact of downsampling on the PSD of pulse energy fluctuations by measuring noise on three photodetected optical signals: a shot-noise-limited telecom-band CW laser, a 10 GHz pulse train generated by passing this laser through cascaded optical phase and intensity modulators[26] and then a low-noise EDFA, and this pulse train after downsampling by a factor of four to 2.5 GHz repetition rate with no additional amplification after downsampling. Shown in Figure 3 are curves for each signal of the fluctuations $\sqrt{S_I(f = 50 \text{ MHz})}$ in the detected photocurrent at a Fourier frequency of 50 MHz versus the total time-averaged detected photocurrent $I$ from the optical signal. To measure the scaling of noise with optical power, these curves are generated by beginning with an optical signal which yields more than 800 μA of detected photocurrent and attenuating this signal before photodetection. The data indicate that both the pulse-generation process and the downsampling process contribute some amount of technical noise at 50 MHz Fourier frequency to the photocurrent, because the measured curves are well-modelled by a quadrature sum of a shot-noise contribution and a technical noise contribution. The contributions of these two types of noise can be determined because they scale differently with the photodetected power: shot noise obeys the relationship $\sqrt{S_I(f = 50 \text{ MHz})} = \sqrt{2e\langle I \rangle}$, $\langle I \rangle$ denoting the time-averaged photocurrent, while the technical-noise contribution arises from fluctuations in the expected photocurrent $I(t)$ and scales linearly with the detected photocurrent. We observe that downsampling by a factor of four leads to a multiplication of the amplitude of the technical noise by a factor of ~1.7 on the optical signal relative to the carrier, which due to finite noise bandwidth is somewhat less than the factor of two (four, in electrical power units) which would be expected for ideal downsampling by a factor of four in the presence of white technical noise. These results further demonstrate that, properly implemented, downsampling does not magnify noise on the pulse train to a degree which is prohibitive for applications.

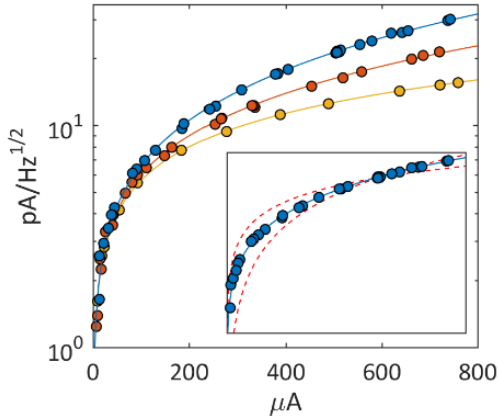

Figure 3. Fluctuations at 50 MHz Fourier frequency in the detected photocurrent as a function of the time-averaged photocurrent in three cases: CW laser at the shot-noise limit (lowest, yellow), 10 GHz pulse train (middle, red), and 2.5 GHz down-sampled pulse train (highest, blue). Dots show measured data and curves show fits to the data. The fit for the shot-noise-limited laser has a single free parameter, which is a scaling factor of order 1 due to frequency dependence of the photo-detector's trans-impedance gain. The fits for the pulse trains have a scaling factor in common, and have as an additional parameter the amplitude of the technical noise on the pulse train. This is -153.9 dBc/Hz for the 10 GHz pulse train and increases by a factor of ~1.7[2] to -149.3 dBc/Hz for the 2.5 GHz downsampled pulse train. Inset: Optimized fits (dashed red) to the experimental data for the downsampled 2.5 GHz pulse train using only shot-noise or linear technical noise scaling, demonstrating that both noise processes are important for explaining the data.

**E. Model for effects of incomplete extinction of rejected pulses and amplification of a downsampled pulse train**

To this point, we have considered effects of downsampling assuming that extinction of the rejected pulses is complete, but in a practical application this is not necessarily the case. The modulators used for pulse extinction may transmit a substantial amount of energy from the rejected pulses—for example, one commercial manufacturer specifies 25 dB extinction ratio, this number varies in practice. Additionally, the electronic gating signal may not have sufficient bandwidth to completely switch from transmission to extinction within the repetition period of the incoming pulse train, and initial extinction can be followed by some transmission caused by ringing in the gating signal. Bandwidth limitations will be increasingly likely as the repetition rates of frequency combs increase, placing more demanding requirements on gating electronics. Incomplete extinction will add modulations to the optical spectrum and will raise the total power of the downsampled pulse train while keeping the energy of the fully-transmitted pulses fixed. This will require higher average power to achieve a given target pulse energy.

The effects of incomplete extinction of rejected pulses are exacerbated if the incomplete extinction does not happen in a deterministic and repetitive fashion; this could occur, for example, if intermediate pulses fall near the edge of the gate in the presence of relative timing jitter between the optical and electronic pulse trains; or if the extinction ratio fluctuates in time. Interestingly, if the downsampled pulse train is subsequently amplified and spectrally broadened, the impact of incomplete extinction depends on whether the optical amplifier used operates in the linear regime or in the saturated regime.

As an example, we consider the case where each fully transmitted pulse is preceded and followed by partially-extinguished pulses whose amplitudes fluctuate for each period of the downsampled pulse train. This fluctuation could occur because the pulses lie on the edge of the electronic gate and there is relative timing jitter between the optical pulse train and the gating signal. It is true that these fluctuations will lead to decoherence during nonlinear spectral broadening. However, the coherence is degraded by this mechanism only within the bandwidth that is achieved by the broadened, partially-extinguished pulses. In efficient *f-2f* interferometry only the fully-transmitted pulses should reach an octave in bandwidth. Therefore, this mechanism of supercontinuum decoherence is not a problem in *f-2f* interferometry in general, unless there is coupling between the amplitudes of the amplified partially-extinguished pulses and the amplified fully transmitted pulses. This coupling can arise, for example, through amplification in saturation, which then leads to decoherence across the full bandwidth of the supercontinuum.

To illustrate this point, we have performed numerical simulations of the spectral broadening of a 100 GHz train of 100 fs pulses which has been downsampled to 10 GHz and then amplified. We use an adaptive[27] split-step Fourier method[28] to simulate spectral broadening in 30 cm of HNLF according to the generalized nonlinear Schrodinger equation[29]. In the simulation each fully-transmitted pulse, amplified to 1 nJ, is preceded and followed by partially-extinguished pulses with normally distributed and uncorrelated energies with mean of 0.3 nJ and standard deviation of 0.225 nJ. This models the effect of adjacent pulses which coincide with the edge of the gate. We simulate amplification in two regimes: saturation is simulated using a fixed-energy method wherein the pulse energies in each three-pulse burst are re-scaled by a common factor so that the total energy is 1.6 nJ; linear amplification is simulated using a fixed-gain model, which involves no such rescaling of pulses. Numerically, we simulate the spectral broadening of each pulse individually, which is acceptable because terms in the generalized non-linear Schrodinger equation operate only locally or, in the case of the Raman term, on the timescale of several femtoseconds, while the separation between the pulses in each burst is 10 ps (the inverse of the initial 100 GHz repetition rate). We have verified that during simulated time-evolution each broadened pulse remains well-centered in its 5 ps simulation window.

Results of this study are shown in Figure 4. Figure 4a depicts a three-pulse burst before and after propagation in HNLF. In Figure 4b we show spectra corresponding to spectral broadening of this three-pulse burst, as well as plots of the spectral coherence averaged over many simulations. The first-order spectral coherence $g_{12}^{(1)}(\lambda)$ is defined as:

$$\left|g_{12}^{(1)}(\lambda)\right| = \left|\frac{\langle E_1^*(\lambda)E_2(\lambda)\rangle}{\sqrt{\langle|E_1(\lambda)|^2\rangle\langle|E_2(\lambda)|^2\rangle}}\right| = \left|\frac{\langle E_1^*(\lambda)E_2(\lambda)\rangle}{\langle|E(\lambda)|^2\rangle}\right|.$$

Curves are plotted for the fixed-gain and fixed-energy cases, as well as for the case with ideal downsampling (no partially-extinguished pulses) and only shot-noise on the pulse train. The averages in the formula above are over 1000 instantiations of the pair $E_1$ and $E_2$, for a total of 2000 broadened spectra for each pulse within the burst of three. In both the fixed-gain and fixed-energy cases the coherence is poor in the center of the spectrum, but in the fixed-gain case, which models amplification in the linear regime, the coherence is preserved in the high- and low-frequency ends of the spectrum where it is needed for self-referencing.

## 3. DISCUSSION

Downsampling via pulse gating is a promising tool to manipulate high-repetition-rate frequency combs from low size, weight, and power packages and to aid in the detection of their offset frequencies. In our experiments downsampling enabled detection of $f_0$ at a signal-to-noise ratio sufficient for measurement and stabilization, which otherwise would have required significantly higher average power. The effects of the electronic timing jitter of the gate signal are negligible so long as incoming optical pulses do not arrive coincidentally with the edge of the gate; when they do, timing jitter induces amplitude noise on the transmitted pulses. This results in an increase in RMS optical pulse energy fluctuations $\sigma_{PEF}$. Independently, the PSD of pulse energy fluctuations may be increased by aliasing of technical noise and by shot noise, depending on the relative magnitudes of these two types of noise. Each of these sources of signal-to-noise-ratio degradation has the potential to interfere with detection of $f_0$. This investigation of these challenges will facilitate application of the technique in high-repetition-rate frequency comb systems. Importantly, our experiments demonstrated that downsampling does not add a significant amount of noise to the frequency components of the pulse train, and in a separate experiment the technique has recently been used successfully to detect the carrier-envelope offset frequency of a 10 GHz comb by downsampling by a factor of four[20].

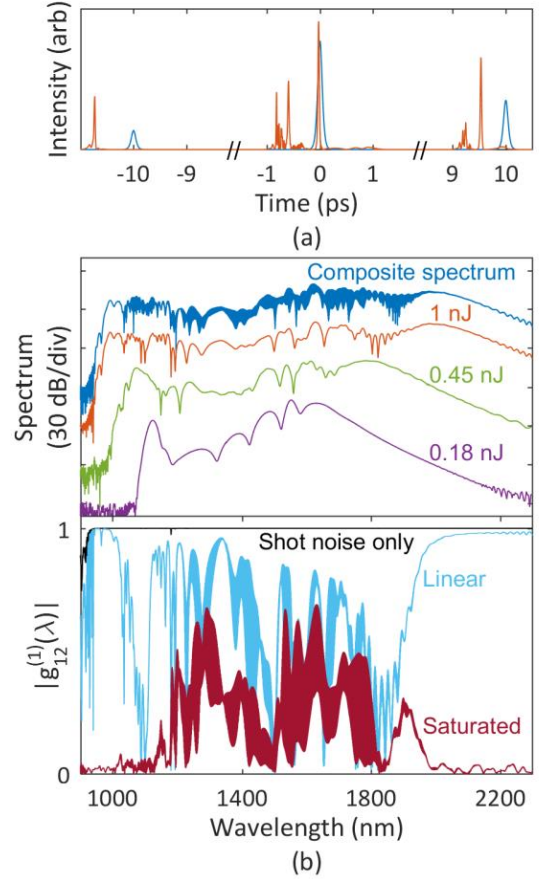

Figure 4. Investigation of incomplete pulse extinction and amplification. a) A burst consisting of a fully-transmitted 1 nJ, 100 fs pulse and 100 fs partially-transmitted adjacent pulses with energies of 0.18 nJ and 0.45 nJ. Blue indicates initial sech$^2$ pulses, and orange indicates the intensity after propagation through 30 cm HNLF. Note that the x-axis has been broken. b) Top panel: optical spectra corresponding to the pulses shown in orange in (a), showing the composite spectrum of the three pulses (top, blue) and the spectra of the 1 nJ central pulse (second, orange), the 0.45 nJ adjacent pulse (third, green), and the 0.18 nJ adjacent pulse (bottom, purple). Bottom panel: Calculated spectral coherence averaged over 2000 simulations for the case of shot-noise only (top, black) and for the case of fluctuating amplitudes of the first and last pulses as described in the text, after simulated amplification in a linear-regime optical amplifier (second, teal), and a saturated optical amplifier (bottom, maroon). For the case of linear-regime operation, high spectral coherence is preserved in the extreme ends of the supercontinuum even as it is lost in the center, in contrast with the complete loss of coherence after amplification in saturation.

To employ downsampling as demonstrated here with repetition rates >10 GHz will require electronic gates with duration ≤100 ps. Technology to downsample with gates as short as 20 ps is commercially available, while 100 Gb/s integrated circuits and 25 GHz demultiplexing have been demonstrated[30,31]. Barring the use of such state-of-the-art electronics, pulse gates of duration longer than the incoming optical pulse train's repetition period can be employed. This will be technically easier to achieve, but will result in additional modulations on the spectrum of the downsampled pulse train.

The ambiguity of the input comb's offset frequency as a result of the reduction of the offset frequency modulo the new repetition rate makes downsampling most suitable for applications where the ambiguity can be removed by some other method. Two such applications are frequency comb calibration of astronomical spectrographs, where measurement of the wavelength of a comb mode can remove the ambiguity, and microresonator-based frequency combs, where the uncertainty in the offset frequency is determined by the frequency stability of the pump laser and can be much less than the repetition rate of the downsampled comb.

**Funding.** This work is supported by the DARPA QuASAR and PULSE programs, NIST, NASA, and the Air Force Office of Scientific Research under award number FA9550-16-1-0016. D.C.C acknowledges support from the NSF GRFP under grant no. DGE 1144083.

**Acknowledgement.** We thank Fabrizio Giorgetta and Lora Nugent-Glandorf for thoughtful comments on this manuscript and M. Hirano of Sumitomo Electric Industries for providing the HNLF.